\documentclass[11pt,twoside]{article}

\usepackage{asp2014}

\aspSuppressVolSlug
\resetcounters

\begin{document}

\title{The International Virtual Observatory Alliance in 2019}


\author{Mark G. Allen,$^1$ Patrick Dowler,$^2$ Janet D. Evans,$^3$ Chenzhou Cui,$^4$ Tim Jenness$^5$, Bruno Mer\'in$^6$, G. Bruce Berriman$^7$ and JJ Kavelaars$^2$ for the IVOA Executive Committee and Technical Coordination Group}
\affil{$^1$Observatoire astronomique de Strasbourg, Universit\'{e} de Strasbourg, CNRS, UMR 7550, Strasbourg, F-67000, France, \email{mark.allen@astro.unistra.fr}}
\affil{$^2$Canadian Astronomy Data Centre, National Research Council Canada, Victoria, British Columbia, Canada}
\affil{$^3$Center for Astrophysics, Harvard \& Smithsonian, Cambridge, MA, USA}
\affil{$^4$National Astronomical Observatories, CAS, Chaoyang District, 100101 Beijing, China}
\affil{$^5$Large Synoptic Survey Telescope, Tucson, AZ, 85719, USA}
\affil{$^6$ European Space Astronomy Centre, Villaneuva de la Ca\~nada, Spain}
\affil{$^7$ Caltech/IPAC, Pasadena, CA~~  91125, USA}


\paperauthor{M. G. Allen}{mark.allen@astro.unistra.fr}{orcid.org/0000-0003-2168-0087}{Observatoire astronomique de Strasbourg, Universit\'{e} de Strasbourg, CNRS, UMR 7550}{}{Strasbourg}{Observatoire Astronomique de Strasbourg}{67000}{France}
\paperauthor{Patrick Dowler}{patrick.dowler@nrc-cnrc.gc.ca}{0000-0001-7011-4589}{Canadian Astronomy Data Centre}{National Research Council Canada}{Victoria}{British Columbia}{V9E 2E7}{Canada}
\paperauthor{Janet D. Evans}{janet@cfa.harvard.edu}{0000-0003-3509-0870}{Center for Astrophysics | Harvard & Smithsonian}{HEAD/CXC}{Cambridge}{MA}{02347}{USA}
\paperauthor{Chenzhou Cui}{ccz@bao.ac.cn}{0000-0002-7456-1826}{National Astronomical Observatories, CAS}{China-VO}{Chaoyang District}{Beijing}{100101}{China}
\paperauthor{Tim~Jenness}{tjenness@lsst.org}{0000-0001-5982-167X}{LSST}{Data Management}{Tucson}{AZ}{85719}{U.S.A.}
\paperauthor{Bruno Merin}{}{0000-0002-8555-3012}{ESAC}{}{Villaneuva de la Ca\~nada}{}{}{Spain}
\paperauthor{G. Bruce~Berriman}{gbb@ipac.caltech.edu}{0000-0001-8388-534X}{Caltech/IPAC}{}{Pasadena}{CA}{91125}{USA}
\paperauthor{JJ Kavelaars}{JJ.Kavelaars@nrc-cnrc.gc.ca}{0000-0001-7032-5255}{Canadian Astronomy Data Centre}{National Research Council of Canada}{Victoria}{British Columbia}{V9E 2E7}{Canada}



\begin{abstract}
The International Virtual Observatory Alliance (IVOA) held its bi-annual Interoperability Meetings in May 2019, and in October 2019 following the ADASS XXIX conference. We provide a brief report on the status of the IVOA and the activities of the Interoperability Meetings.
\end{abstract}



\section{Introduction}


The International Virtual Observatory Alliance\footnote{\url{http://www.ivoa.net}} (IVOA) develops  the technical standards needed for seamless discovery and access to astronomy data worldwide, according to the FAIR principles \citep{doi:10.1038/sdata.2016.18}, thus realizing the Virtual Observatory (VO).
IVOA is a framework for discussing and sharing  ideas and technology, for engaging astronomy projects, missions and researchers, and for promoting and publicising the VO.
The VO is now a mature framework for the interoperability of astronomy data, with IVOA compliant services operated by astronomy data centers worldwide.
The IVOA work is pursued by Working Groups (WG) and Interest Groups (IG)  coordinated by the Technical Coordination Group (TCG), guided by a scientific priorities committee (CSP), with the overall direction provided by the IVOA Executive Committee. 

The IVOA holds bi-annual Interoperability Meetings which are focal points for the progress of its activities.
In 2019 these {\em Interop} meetings have been held in: May in Paris, France\footnote{https://wiki.ivoa.net/twiki/bin/view/IVOA/InterOpMay2019}, with 131 attendees; and in October following the ADASS XXIX conference in Groningen, Netherlands\footnote{https://wiki.ivoa.net/twiki/bin/view/IVOA/InterOpOct2019}, with 125 attendees.
This article provides an overview of some of the topics being addressed by IVOA in 2019.
A report of the IVOA activities at the 2018 November meeting is available in \citet{H2_adassxxviii}.
 
\section{IVOA Interoperability Meetings in 2019}

A session on "Big Data Challenges in Astronomy" in May brought current and future projects together to outline their data exploitation plans and to participate in a larger discussion on how the IVOA infrastructure and protocols can help to face these challenges. The participating projects were LSST, the ESA Gaia and Euclid missions, Pan-STARRS, WFIRST, TESS, and SKA Regional Centres. 


The discussion of the various plans included a comparison of the expected maximum volumes of data, data distribution mechanisms, needs for "code to the data" and identification of relevant IVOA standards. The data volumes ranged from $\sim$2PB (Pan-STARRS and Gaia), to 10s of PB for LSST and Euclid, and 600 PB/year for SKA. The planned data access mechanisms emphasized evolution towards science platforms that will employ Jupyter notebooks with \texttt{astroquery}-compliant interfaces deployed on commercial and institutional clouds. IVOA standards and infrastructure components were identified as important for all of these projects, with TAP+ADQL, SODA, DataLink and VOSpace being the most mentioned. The feedback obtained helped to identify things that are currently missing or in need of further development such as re-binning and re-sampling in SODA cut-out services, interoperable notebooks, and standards for "code-to-the-data".


A special joint event of IVOA and Astropy was held during the May meeting in the form of an experimental VO-Astropy sprint with the goal of adding or enhancing VO tools within Astropy and supporting a higher level of interaction between IVOA and Astropy. It was decided to resurrect development of PyVO and move it directly into the Astropy organization on GitHub. This has resulted in renewed interest with many contributions being accepted from LSST, CADC, and others.

Further engagement with radio astronomy projects was pursued in October with a session on ``Radio Astronomy in the VO'' to discuss ways in which the IVOA standards can evolve to meet their use cases. Emphasis was put on the immediate improvements in discovering radio astronomy data in the VO, such as standard descriptions for UV plane coverage, antennae configurations, largest angular scales and beam/PSF characterisation.

\subsection{Working Group and Interest Group Summaries}

\noindent\textbf{Applications WG:}
The applications sessions included many demos involving TOPCAT, CASSIS, WWT, ESASky and pyESASky, Aladin, MOCpy, and PyVO. Standards discussions mainly involved spacetime MOC \citep{O2-3_adassxxix}, Web SAMP over \texttt{https} \citep{P2-7_adassxxix}, and VOTable.

\noindent\textbf{Data Access Layer WG:}
One of the new areas being considered in this group is standards for coordination of multi-messenger follow-up observation campaigns, in particular to share information on the “visibility” of sky regions and a protocol to query such information. These are being developed as the ObsLocTAP and ObjVisSAP standards. 
There were also discussions on how to handle complex hierarchical results from TAP queries \citep{P2-15_adassxxix}.

\noindent\textbf{Data Modeling WG:}
The DM group is working on various models including models for coordinates, time series and provenance \citep[see e.g.,][]{P2-6_adassxxix}, and their expression in VO-DML. 
A long term effort has been pursued on the difficult task of developing models to describe Space-Time-Coordinates. This has been re-organised into separate models for coordinates, transformations and measurements data models. Other topics discussed include the Common Archive Observation Model and how it fits into the IVOA model framework, and also a model for astronomical sources in catalogues.

\noindent\textbf{Grid and Web Services WG:}
The main focus of the GWS session was the situation with authentication.
There still remains an open debate as to whether OAuth2 is suitable in all cases (given that it requires a web browser to be involved) and whether cookies or tokens are to be preferred. Science platforms continue to be popular, all using different interfaces and deployment mechanisms with some desire to look for commonalities.

\noindent\textbf{Registry WG:}
The connection of VO registries to other data infrastructures was pursued, in particular via the participation of the EUDAT project (in particular for the B2FIND registry).

\noindent\textbf{Semantics WG:}
The use and standardisation of vocabularies was a topic of Semantics IG sessions, and this included discussion of the relation of VO vocabularies to the Unified Astronomy Thesaurus. The need for a telescope and instrument index using a standard nomenclature was identified. The process for the maintenance of Unified Content Descriptor semantic tags was updated in 2019.

\noindent\textbf{Data Curation and Preservation IG:}
The DCP interest group discussed the current status of the DOI Note that is being planned to describe best practices for DOI usage for astronomy data
\citep[see also][]{P2-19_adassxxix}.
There was also a discussion on reproducibility and on the future need to specify usage licenses along with datasets.

\noindent\textbf{Education IG:}
Education activities in IVOA covered many different aspects of VO tools and services being used in scientific tutorials for astronomy researchers, and also for school education, citizen science, and outreach purposes. The sessions have also included the participation of the International Planetarium Society in the May meeting, and the October meeting was held in the DotLivePlanetarium centre with the benefit of spectacular projections of data onto the dome, stimulating the ideas (and access mechanisms and formats) for use of VO data for wider public and education audiences.

\noindent\textbf{Knowledge Discovery IG:}
Is the VO ready for data science? -- this was a topic of this interest group, with lively discussion on the use of VO data access for use in machine learning tools, and the potential for innovative uses of the data.

\noindent\textbf{Operations IG}
The Ops group sessions have included reports on the various efforts to monitor VO services in terms of validation and reliability. The centralisation of information about the available validators has helped these to be more widely used for a wide range of the IVOA standards. 

\noindent\textbf{Solar System IG:}
Members of the International Planetary Data Alliance (IPDA) attended the May meeting which facilitated interaction between IVOA and IPDA on common interoperability issues. VO for solar physics was also a topic of the May meeting. A contribution on “Sharing Solar data with ESCAPE and SOLARNET” reviewed the current status of sharing space- and ground-based solar observation data, and the various levels of standardisation and usage of the Solar Virtual Observatory. Discussion focused on TAP (and its customisation EPN-TAP), SAMP for tool interoperability, UCDs for semantics with new requirements for use of UCDs in VOEvents. 

\noindent\textbf{Time Domain IG:}
The VOTable standard has been upgraded to include the TIMESYS element which is a simple way of including time metadata, and this has been demonstrated in VizieR and GAVO services, with client implementations in Topcat, Stilts and SPLAT. Another topic that was pursued in coordination with the Apps group is the connecting of data "in space and time" to enable searches for data based on "where and when" using a mechanism that builds on the MOC sky coverage, adding the time dimension to become ST-MOC. Various examples were shown based on the use case of finding the data taken during simultaneous observations from different missions. 

\section{Standards Development Process}

Standards development follows a well described process \citep{2017ivoa.spec.0517G} but has mostly taken place using mailing lists, a Subversion repository or Word documents, and a Wiki.
Over the past year we have decided to begin migrating away from this tooling and will soon begin to develop standards using GitHub, developing each standard in its own git repository, tracking change requests using GitHub issues, and using Pull Requests to discuss suggested edits.
The \texttt{master} branch will always show the current development version and we will use tags to indicate the different stages of the release process.
The standards are hosted at \url{https://github.com/ivoa-std/}.



\acknowledgements The IVOA would like to thank the local organisers of the IVOA Interoperability Meetings in Paris and Groningen, and the ADASS XXIX conference. 

\bibliography{P11-13}  


\end{document}